# Study of high-energy particle acceleration in Tycho with gamma-ray observations

**Nahee Park* for the VERITAS Collaboration†**
*The University of Chicago*
*E-mail:* nahee@uchicago.edu

Gamma-ray emission from supernova remnants (SNRs) can provide a unique window to observe the cosmic-ray acceleration believed to take place in these objects. Tycho is an especially good target for investigating hadronic cosmic-ray acceleration and interactions because it is a young type Ia SNR that is well studied in other wavelengths, and it is located in a relatively clean environment. Several different theoretical models have been advanced to explain the broadband spectral energy emission of Tycho from radio to the gamma-ray emission detected by the *Fermi*-LAT in the GeV and by VERITAS in the TeV.

We will present an update on the high-energy gamma-ray studies of Tycho with $\sim$ 150 hours of VERITAS and $\sim$ 77 months of the *Fermi*-LAT observations, which represents about a factor of two increase in exposure over previously published data. VERITAS data also include exposure with an upgraded camera, which made it possible to extend the TeV measurements toward lower energy, thanks to its improved low energy sensitivity. We will interpret these observations in the context of the particle acceleration in Tycho and proposed emission models.

*Speaker.
†veritas.sao.arizona.edu

## 1. Supernova Remnant Tycho as a cosmic ray accelerator

Supernova Remnants (SNRs) are broadly accepted as the main accelerators of galactic cosmic rays (CRs) with energies up to the knee region ($\sim 10^{15}$ eV)([1, 2]). Observations of X-ray synchrotron emission from SNRs provide clear evidence for the acceleration of electrons to energies of about 100 TeV, but similar evidence for proton acceleration is lacking. Recent measurements of pion-decay signatures from W44 and IC 443 by the *Fermi*-LAT provide indirect evidence of the acceleration of hadronic CRs in SNRs [3]. However, the maximum energy to which SNRs can accelerate hadronic CRs remains unknown. Observations of SNRs in the gamma-ray energy range can provide a unique window to study the high energy hadronic acceleration in SNRs, but that requires discriminating between possible leptonic and hadronic origins of the gamma-ray emission. The interpretation of gamma-ray emission from SNRs is difficult because many of the relevant quantities are unknown. The presence of data at other wavelengths can help significantly in this regard.

Tycho is a good SNR for the study of particle acceleration because (1) it has been widely observed in the IR, optical, radio and X-ray; (2) it is a young, Type Ia SNR, so we do not expect the progenitor star to have modified the environment as in core-collapse supernovae; and (3) it sits in a relatively clean environment. Thus, it forms a good candidate for high-energy observations and detailed modelings. While the acceleration of electrons in the remnant was indirectly shown by the measurement of non-thermal hard X-ray emissions by *Chandra* [4] and *NuSTAR* [5], detection of gamma-ray emissions at GeV [6] and TeV [7] energies provided data to study the acceleration of hadrons in Tycho. Several models were developed to explain the full spectral measurements of Tycho's SNR from radio to gamma-ray including the morphology of radio and X-ray emissions. Many of these proposed models agree that hadronic emission is the dominant component of the gamma-ray emission from Tycho. However, the details of the mechanisms for the emission vary considerably. VERITAS has pursued a deep observation on Tycho to provide improved gamma-ray measurements of the remnant. In this paper, we will present an update on the high-energy gamma-ray studies of Tycho by using the *Fermi*-LAT and VERITAS data with a factor of two increased exposure over previously published data.

## 2. Gamma-ray observations and data analysis

### 2.1 *Fermi* observation and analysis

The *Fermi* Large Area Telescope (*Fermi*-LAT) was launched in 2008 June, and has provided all sky coverage in the GeV energy range over seven years. Pass 7 reprocessed data from the *Fermi*-LAT were used to study the GeV emission from Tycho. We analyzed a data set of $\sim 77$ months, selecting events with energies from 100 MeV to 300 GeV that fall within a radius of $30°$ centered on the position of the remnant. The publicly available Fermi Science Tools[1] were used for the analysis. For checking the systematics of the spectral reconstruction, two different instrument response functions (IRFs), *P7REP_SOURCE_V*15 and *P7REP_CLEAN_V*15, were used. For the image of the remnant, analysis with front conversion of *P7REP_SOURCE_V*15 class was also checked. All recommended quality cuts for standard analysis were implemented. All sources from the 3FGL catalogue which fall within the radius $40°$ around Tycho were included in the analysis. Binned likelihood analysis was performed first on the 3FGL source associated with Tycho ($3FGLJ0025.7+6404$), and the results were later checked with analysis on the best source position estimated from the test statistic (TS) map. The spectra of sources from the 3FGL catalogue were used for the analysis. Normalization values for the diffuse background components as well as sources that are located within a radius of $10°$ around Tycho were freed for the likelihood analysis.

---

[1] http://fermi.gsfc.nasa.gov/ssc

## 2.2 VERITAS observation and analysis

VERITAS is an array of four atmospheric Cherenkov telescopes designed to study astrophysical sources of gamma-ray emission in the 85 GeV-30 TeV range. Located at the Fred Lawrence Whipple Observatory in southern Arizona [8], VERITAS has the sensitivity to detect a point source with a flux of 1% of the Crab Nebula flux within 25 hours with an angular resolution better than 0.1° at 1 TeV. VERITAS has observed Tycho's SNR since 2008, collecting a total of ∼ 150 hours of data over five years. This deep observation of Tycho overlaps with two major upgrades of VERITAS: the relocation of telescope 1 in 2009 and a camera upgrade in 2012. The performance of VERITAS has been improved with these upgrades [9]. The discovery of gamma-ray emission from Tycho was reported by VERITAS based on 67 hours of observation during 2008-2010. As shown in Table 1, VERITAS has accumulated a data set more than twice as large since the discovery. In particular, 74 hours of observation were collected with the upgraded array from 2012 with enhanced sensitivity at energies lower than a few TeV.

| Before telescope 1 relocation 2008-2009 | After telescope 1 relocation (2009-2011) | After camera upgrade (2012-2015) |
|---|---|---|
| ∼ 22 hours | ∼ 55 hours | ∼ 74 hours |

Table 1: VERITAS observations on Tycho after quality cuts. Good-weather data taken with four active telescopes were used for this study.

For the analysis, cuts were selected *a priori* to get a good sensitivity for a source with 0.9% strength of the Crab Nebula. Cuts were optimized separately for the 2009-2011 data set and for the 2012-2015 data set. Optimized cuts for 2009-2011 were used for the 2008-2009 data set. During the optimization process, differential sensitivities were checked to gain reasonable sensitivity with a low energy threshold. As a result, cuts for the 2009-2011 data set have an energy threshold value of ∼ 800 GeV, similar to the analysis of the discovery paper, while cuts for the 2012-2015 set have a lower energy threshold value of ∼ 400 GeV for Tycho's average elevation angle of 55° without hurting the overall sensitivity.

## 3. Results

### 3.1 The *Fermi*-LAT results

The overall fit from 100 MeV to 300 GeV yields a TS value of 32 for Tycho, assuming a power-law distribution. The measured integral flux is $(5.7 \pm 2.8_{stat} \pm 1.4_{sys}) \times 10^{-9} cm^{-2} s^{-1}$ with a gamma index of $1.97 \pm 0.15_{stat} \pm 0.1_{sys}$. The results agree within $1\sigma$ with the discovery paper [6]. The TS value reported in the discovery paper was 33, which is comparable to the results even though more than twice as many data were used in this analysis. This is mainly because of the low energy range (100 MeV - 300 MeV) included in this study. The analysis with an energy range of 300 MeV - 300 GeV showed a TS value of 52 with a spectral index of $2.3 \pm 0.2_{stat} \pm 0.1_{sys}$.

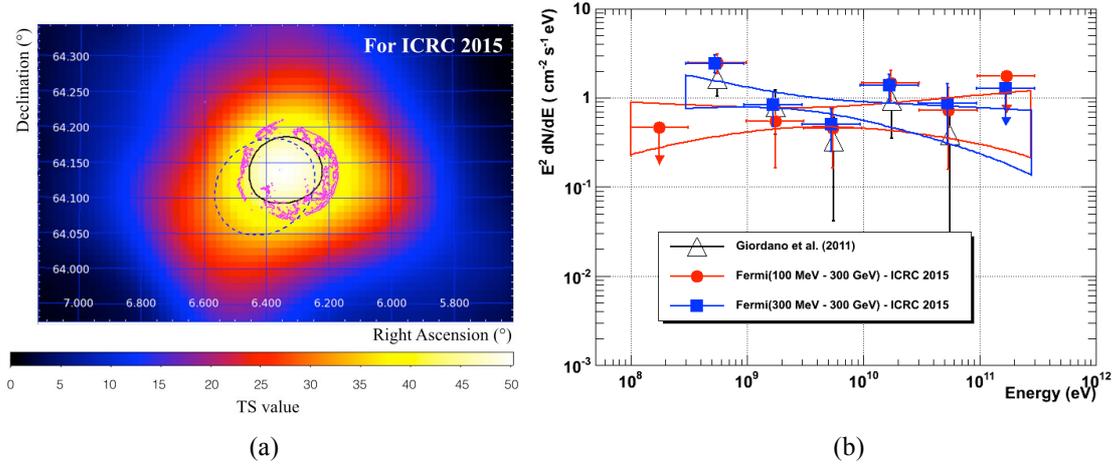

(a)  (b)

Figure 1: *(a):* *Fermi* TS map with *P7REP_CLEAN_V*15 IRF for energy higher than 1 GeV. The magenta contour is the *Chandra* X-ray emission with energy higher than 4.1 keV[2]. The blue dotted line is the previously published 95% confidence area for the *Fermi*-LAT position. The black line is the 95% confidence area for energy higher than 1GeV from this study, defined by the ΔTS for 2 degrees of freedom. *(b):* Fermi spectrum with *P7REP_SOURCE_V*15 IRF with energy range between 100 MeV and 300 GeV compared with analysis done with energy range between 300 MeV and 300 GeV.

Figure 1a shows a TS map generated using the *Fermi* tool *gttsmap* for the region around Tycho. The TS maps for three different IRFs were checked for energy higher than 1 GeV, and they all agree very well. The 95% confidence area from this study matches well with the 95% confidence area from the previous paper, as well as with the position of the 3FGL source associated with Tycho.

The entire energy range was divided into evenly spaced energy bins in logarithmic scale to get the spectral energy distribution (SED). An individual likelihood analysis was performed for each bin using the fitting result from the analysis of entire energy range. All parameters except the normalization of Tycho were fixed. The flux was calculated for bins with positive TS values while the 95% upper limit value was calculated for bins with TS value less than or equal to zero. Figure 1b shows the SED with the *P7REP_SOURCE_V*15 IRF for two different energy ranges compared to the published result. Only statistical errors are shown in the figure. Calculation with the other IRF (*P7REP_CLEAN_V*15) showed a comparable result.

Intrigued by the low TS value in 100 MeV - 300 MeV energy bin, we tested a broken power-law model as well. Based on the TS value difference in the power-law and broken power-law models, the latter is preferred at the level of $2\sigma$.

---

[2] http://chandra.harvard.edu

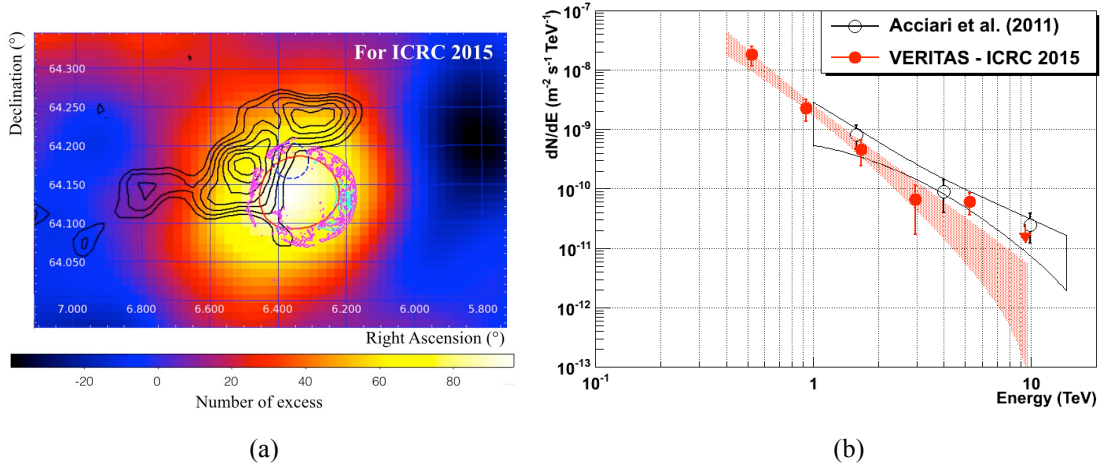

(a)  (b)

Figure 2: *(a):* Smoothed VERITAS gamma-ray count map of the region around Tycho's SNR. The 1$\sigma$ statistical error on the published centroid position is drawn with a dotted circle. Black contours are the $^{12}CO$ (J=1-0) emission from the FCRAO Survey [10]. *Chandra*'s measurement of the X-ray emission with energies larger than 4.1 keV is shown by the magenta contours. *NuSTAR*'s measurements of X-rays in the energy range between 20 keV and 40 keV after smoothing is shown by the cyan contours. The 95% confidence area for energies higher than 1 GeV from the *Fermi* analysis explained in Section 3.1 is shown by the red contours. *(b):* VERITAS spectra. The previous result is plotted as black empty circles and the result of the present study is shown with filled red circles. Flux errors were calculated from error propagation of the fitting function and drawn as an area around the data points.

### 3.2 VERITAS results

The analysis with the combined set of VERITAS data detected gamma-ray emission from Tycho with a post-trials significance of 6.3$\sigma$. Figure 2a shows the smoothed excess map with the previously published centroid position and statistical uncertainty of the location. Figure 2b shows the spectrum analysis of this study in comparison to the previous result. With the enhanced sensitivity of the upgraded camera, the VERITAS measurements have been extended down to $\sim$ 400 GeV. The spectrum is consistent with a power-law $dN/dE = N_0(E/1TeV)^{-\Gamma}$ with a normalization $N_0$ of $(2.2 \pm 0.5_{stat}) \times 10^{-13} cm^{-2} s^{-1} TeV^{-1}$ and a gamma index $\Gamma$ of $2.92 \pm 0.42_{stat}$. The reduced chi-square of the fit is 1.34 (4.01/3). Above 7.5 TeV, the gamma-ray excess has a significance of $\sim 1\sigma$, and a 99% upper limit of $2.5 \times 10^{-15} cm^{-2} s^{-1} TeV^{-1}$ was obtained by Rolke's method [11] calculated with an index of 2.9.

Previous results reported a spectrum consistent with a power-law distribution with gamma index of $1.95 \pm 0.51_{stat} \pm 0.30_{sys}$. While the index of the combined data set is softer than this, the results match within errors. Subsets of the data were selected to check the consistency between epochs and with the previous published data set. This investigation has revealed that the extension to lower energies, which was enabled by the VERITAS upgrades, is primarily responsible for driving the index to softer values.

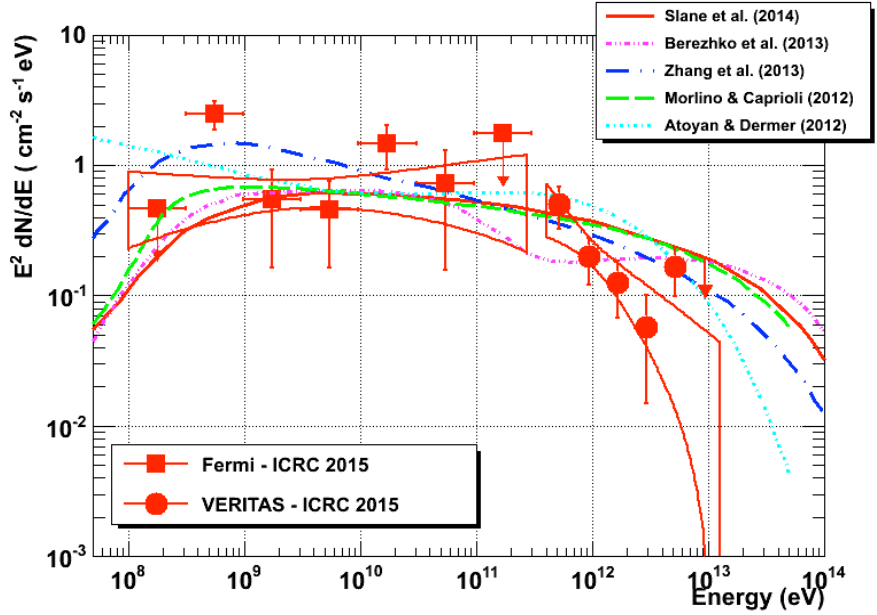

Figure 3: Fermi and VERITAS SEDs with theoretical models. Filled red squares show the *Fermi* results and filled red circles show the VERITAS results from this study. Models are shown for Slane et al. [12]'s preferable model A in solid red line, for Berezhko et al. [13]'s model in magenta, for Zhang et al. [14]'s model in blue, for Morlino & Caprioli [15]'s model in green and for Atoyan & Dermer [16]'s leptonic model in cyan.

## 4. Discussion

Figure 2a shows the high energy gamma-ray count map of VERITAS in the Tycho region overlaid with measurements from other wavelengths. The gamma-ray count map has been smoothed with a Gaussian kernel with a radius of 0.06°. Black contours are 115 GHz line emissions associated with the $^{12}CO$ (J=1-0) molecular transition as measured by the 14 m telescope of the Five College Radio Astronomy Observatory (FCRAO) [10], integrated over the velocity range from $-68\,kms^{-1}$ to $-50\,kms^{-1}$. X-ray emission with energy higher than 4.1 keV measured by *Chandra* is shown with the magenta contours, while a recently reported X-ray measurement in the energy range between 20 keV and 40 keV by *NuSTAR* [5] is shown with the cyan contours.

Previous measurements from VERITAS reported a slight displacement toward the northeast of the remnant where a CO cloud exists. Although the displacement was not statistically significant and there is no strong evidence of SNR interaction with the molecular cloud [17], it is still an interesting environment for gamma-ray studies because the cloud can provide target material for escaping cosmic rays. If there is a large contribution from escaping cosmic rays, one would expect to see the centroid change depending on the energy. The morphology change from the GeV energy range to the TeV energy range in W28 can be explained as an example of such case [18]. However, the gamma-ray images of Tycho with *Fermi* and VERITAS look quite similar as shown in Figure 2a. Also, when we compare images of VERITAS data divided into two energy bins (one with energies lower than 800 GeV and the other with energies higher than 800 GeV), we find no significant centroid shift toward the molecular cloud.

Several theoretical models have been developed to explain the gamma-ray emission from Tycho ([12, 13, 14, 15, 16]). Figure 3 shows the updated gamma-ray SEDs overlaid with the existing theoretical models. While statistical tests are needed in order to evaluate them, the models disagree somewhat with the updated gamma-ray measurements. For instance, the softness of the gamma-ray emission for energies higher than 500 GeV produces tension with all of the models.

## 5. Summary

Updated high-energy gamma-ray studies of Tycho were presented with a factor of two increased exposure for both VERITAS and *Fermi*-LAT data. The deeper exposure as well as improved low energy sensitivity allowed us to extend the TeV measurements toward lower energies, which led to somewhat different results than previously expected. The implications of these new results in the context of particle acceleration in Tycho will be discussed during the conference.

## 6. Acknowledgements

This research is supported by grants from the U.S. Department of Energy Office of Science, the U.S. National Science Foundation and the Smithsonian Institution, and by NSERC in Canada. We acknowledge the excellent work of the technical support staff at the Fred Lawrence Whipple Observatory and at the collaborating institutions in the construction and operation of the instrument. The VERITAS Collaboration is grateful to Trevor Weekes for his seminal contributions and leadership in the field of VHE gamma-ray astrophysics, which made this study possible.